\documentclass[a4paper,11pt]{article}
\pdfoutput=1
\usepackage{jcappub}

\usepackage{amssymb,amsmath}
\usepackage{graphics,epsfig}

\newcommand{\be}{\begin{equation}}
\newcommand{\ee}{\end{equation}}
\newcommand{\beq}{\begin{equation}}
\newcommand{\eeq}{\end{equation}}
\newcommand{\bea}{\begin{eqnarray}}
\newcommand{\eea}{\end{eqnarray}}
\newcommand{\ba}{\begin{eqnarray}}
\newcommand{\ea}{\end{eqnarray}}

\def\one{\mbox{1 \kern-.59em {\rm l}}}

\title{\boldmath Constraints on NonCommutative Spectral Action from
  Gravity Probe B and Torsion Balance Experiments}

\vspace{1cm}
\author[a,b,c]{Gaetano Lambiase,}
\author[d]{Mairi Sakellariadou,}
\author[a,b]{Antonio Stabile}

\affiliation[a]{Dipartimento di Fisica ``E.\ R.\ Caianiello",
  Universit\'a di Salerno, 84084 Fisciano (SA), Italy}
\affiliation[b]{INFN, Sezione di Napoli Italy}
\affiliation[c]{International Institute for Advanced Scientific
  Studies, 89019 Vietri sul Mare (SA), Italy}
\affiliation[d]{Department of Physics, King's College London,
  University of London, Strand WC2R 2LS, London, United Kingdom}

\emailAdd{lambiase@sa.infn.it}
\emailAdd{mairi.sakellariadou@kcl.ac.uk}
\emailAdd{astabile@gmail.com}
\date{\today}

\abstract {Noncommutative spectral geometry offers a purely geometric
  explanation for the standard model of strong and electroweak
  interactions, including a geometric explanation for the origin of
  the Higgs field. Within this framework, the gravitational, the
  electroweak and the strong forces are all described as purely
  gravitational forces on a unified noncommutative space-time.  In this
  study, we infer a constraint on one of the three free parameters of
  the model, namely the one characterising the coupling constants at
  unification, by linearising the field equations in the limit of weak
  gravitational fields generated by a rotating gravitational source,
  and by making use of recent experimental data.  In particular, using
  data obtained by Gravity Probe B, we set a lower bound on the Weyl
  term appearing in the noncommutative spectral action, namely $\beta
  \gtrsim 10^{-6}$m$^{-1}$. This constraint becomes stronger once we
  use results from torsion balance experiments, leading to $\beta
  \gtrsim 10^{4}$m$^{-1}$.  The latter is much stronger than any
  constraint imposed so far to curvature squared terms.}

\begin{document}
\begin{flushleft}
KCL-PH-TH/2013-7
\end{flushleft}

\maketitle
\flushbottom
\section{Introduction: Elements of Noncommutative Spectral Geometry}

One of the main quests in theoretical particle physics is the
unification of all interactions, including gravity. While at low
energy scales one can consider an effective theory with physics being
described by the sum of the Einstein-Hilbert action --- based on
diffeomorphism invariance --- and the Standard Model (SM) action ---
based upon internal symmetries of a gauge group --- this is no longer
valid at high energy scales.  As one approaches the Planck energy
scale, the quantum nature of space-time reveals itself and the correct
formulation of geometry should be within a quantum framework. In
constructing a quantum theory of gravity coupled to matter, we will
adopt the philosophy that the interaction between gravity and matter
is the most important aspect of the whole dynamics.  One may speculate that at very
high energy scales quantum gravity will enforce a wildly
noncommutative space-time, while at intermediate scales --- close but below
the Planck scale --- the algebra of coordinates may be assumed as
moderately noncommutative, and if appropriately chosen it can lead to
a purely geometric explanation of the SM coupled to
gravity~\cite{ccm}.

 NonCommutative Spectral Geometry (NCSG)~\cite{ncg-book1, ncg-book2}
 proposes that the SM fields and gravity are packaged into geometry
 and matter on a Kaluza-Klein noncommutative space. Its main goal is
 to unfold the small-scale structure of space-time from our knowledge
 at the electroweak scale; in that sense NCSG follows a bottom-up
 approach complementary to the top-down approach of string theory.
The Standard Model of strong and electroweak interactions is
considered, within the NCSG framework, as a phenomenological model
which will dictate the structure of space-time. According to this
proposal, a few orders of magnitude below the Planck energy scale,
geometry is composed by a two-sheeted space, made from the product of
a four-dimensional compact Riemannian manifold ${\cal M}$ (with a
fixed spin structure) --- describing the geometry of space-time --- and
a discrete noncommutative space ${\cal F}$ --- describing the internal
space of the particle physics model.  Hence, gravity and the SM fields
are put together into matter and geometry on a noncommutative space
made from the product ${\cal M}\times {\cal F}$.  Such a product
space, seen as a four-dimensional internal Kaluza-Klein space
attached to each point with the fifth dimension being a discrete
zero-dimensional space, leads to an {\sl almost commutative}
geometry. Its physical interpretation is that left- and right-handed
fermions are placed on two different sheets with the Higgs fields
being the gauge fields in the discrete dimensions; the
Higgs can be seen as the difference (thickness) between the two sheets.

The choice of a two-sheet geometry --- an almost commutative manifold ---
has a deep physical meaning. As it has been
highlighted in Ref.~\cite{Sakellariadou:2011wv}, this structure is
essential in order to accommodate the gauge symmetries of the SM,
while in addition it incorporates the seeds of
quantisation~\cite{Sakellariadou:2011wv}. More recently, it has been
also shown~\cite{Gargiulo:2013bla} that this structure can account for neutrino
mixing~\cite{ccm,ac2006}.

The noncommutative nature of ${\cal F}$ is encoded in the spectral
triple $\left({\cal A}_{\cal F},{\cal H}_{\cal F}, D_{\cal F}\right)$.
The algebra ${\cal A}_{\cal F}=C^\infty({\cal M})$ of smooth functions
on ${\cal M}$ is an involution of operators on the finite-dimensional
Hilbert space ${\cal H_F}$ of Euclidean fermions; it plays the r\^ole
of the algebra of coordinates.  The operator $D_{\cal F}$ is the Dirac
operator ${\partial\hspace{-5pt}\slash}_{\cal
  M}=\sqrt{-1}\gamma^\mu\nabla_\mu^s$ on the spin manifold ${\cal M}$;
it corresponds to the inverse of the Euclidean propagator of fermions
and is given by the Yukawa coupling matrix which encodes the masses of
the elementary fermions and the Kobayashi--Maskawa mixing parameters.
The operator $D_{\cal F}$ is such that $JD_{\cal F}=\epsilon 'D_{\cal
  F}J$, where $J$ is an anti-linear isometry of the finite dimensional
Hilbert space with
\be
J^2=\epsilon \ \ ,\ \ J\gamma=\epsilon '' \gamma J~;\nonumber
\ee
$\gamma$ is the chirality operator and $\epsilon, \epsilon ', \epsilon
''\in \{\pm 1\}$.  The internal space ${\cal F}$ has dimension $6$ to
allow fermions to be simultaneously Weyl and chiral, while it is
discrete to avoid the infinite tower of massive particles that are
produced in string theory.

To get the SM, the choice of the algebra should be such that it can
account for massive neutrinos and neutrino oscillations --- thus it
cannot be left-right symmetric --- while noncommutative geometry
imposes constraints on the algebras of operators in Hilbert space; in
addition one should avoid fermion doubling. These considerations lead
to the algebra~\cite{Chamseddine:2007ia}
\be
{\cal A}_{\cal F}=M_{a}(\mathbb{H})\oplus M_{k}(\mathbb{C})~,
\ee
with $k=2a$; $\mathbb{H}$ is the algebra of quaternions, which encodes
the noncommutativity of the manifold.  The first possible value for
$k$ is 2, corresponding to a Hilbert space of four fermions. This
choice is however ruled out from the existence of quarks. The next
possible value is $k=4$ leading to the correct number of $k^2=16$
fermions in each of the three generations. Note that the number of
generations is a physical input in the theory. Let us emphasise that
the choice of ${\cal A_F}$ is the underlying input which determines
the physical implications of the model, in particular the particle
content of the theory. In Ref.~\cite{ccm} it has been chosen so that
it leads to the Standard Model of particle physics.

The spectral geometry in the product ${\cal M}\times {\cal F}$  is
given by the product rules:
\begin{eqnarray}
{\cal A} &=& C^\infty({\cal M})\oplus{\cal A_F}\ , \nonumber\\
  {\cal H}&=&L^2({\cal M},S)\oplus{\cal H_F}\ , \nonumber\\
{\cal D}&=&{\cal D_M}\oplus1+\gamma_5\oplus{\cal D_F}~,
\end{eqnarray}
where
$L^2({\cal M}, S)$ is the Hilbert space of $L^2$ spinors and ${\cal
D_M}$ is the Dirac operator of the Levi-Civita spin connection on
${\cal M}$.

To obtain the NCSG action one applies the spectral action principle to
the product geometry ${\cal M}\times {\cal F}$.  The bare bosonic
Euclidean action is
\be {\rm Tr}(f(D_ A/\Lambda))~, \ee
where $D_A=D +A+\epsilon'J AJ^{-1}$ (with $A$ a self-adjoint operator $A=A^\star$ of the form $A=\sum\limits_{j} a_j[D,b_j], \ a_j,b_j\in {\cal A}$) are
uni-modular inner fluctuations, $f$ is a cutoff function and $\Lambda$
fixes the energy scale; we include the eigenvalues of the Dirac
operator that are smaller than the cutoff scale $\Lambda$. This action
can be seen {\sl \`a la} Wilson as the bare action at the mass scale
$\Lambda$.  To obtain the full action functional one has to include
the fermionic part
\be
(1/2)\langle J\psi,D\psi\rangle~,\nonumber
\ee
where $J$ is the real structure on the spectral
triple and $\psi$ is a spinor in the Hilbert space ${\cal H}$ of the
quarks and leptons.

We will concentrate on the bosonic part of the action. Using heat
kernel methods, the trace ${\rm Tr}(f({\cal D}_A/\Lambda))$ can be
written in terms of the geometrical Seeley-De Witt coefficients $a_n$
--- known for any second order elliptic differential operator --- as
$\sum\limits_{n=0}^\infty F_{4-n}\Lambda^{4-n}a_n$; the function $F$ is
defined such that $F({\cal D}_A^2)=f({\cal D}_A)$.  Hence, the bosonic
part of the spectral action expanded in powers of $\Lambda$
reads~\cite{Chamseddine:1996zu}
\begin{equation}
\label{eq:sp-act}
{\rm Tr}\left(f\left(\frac{{\cal D}_A}{\Lambda}\right)\right)\sim
\sum_{k\in {\rm DimSp}} f_{k} \Lambda^k{\int\!\!\!\!\!\!-} |{\cal D}_A|^{-k} + f(0) \zeta_{{\cal D}_A(0)}+ {\cal O}(1)\,,
\end{equation}
with $f_k$ the momenta of the smooth even test (cutoff) function which decays fast at infinity:
\bea\nonumber
\label{eq:moments0}
f_0 &\equiv& f(0)~, \nonumber\\
f_k &\equiv&\int_0^\infty f(u) u^{k-1}{\rm
  d}u\ \ ,\ \ \mbox{for}\ \ k>0 ~,\nonumber\\ \mbox
    f_{-2k}&=&(-1)^k\frac{k!}{(2k)!} f^{(2k)}(0)~.  \nonumber
\eea
In Eq.~(\ref{eq:sp-act}) above, the noncommutative integration is
defined in terms of residues of zeta functions $\zeta_{{\cal D}_A} (s)
= {\rm Tr}(|{{\cal D}_A}|^{-s})$ at poles of the zeta function and the
sum is over points in the dimension spectrum of the spectral triple.

For a four-dimensional Riemannian geometry, the ${\rm Tr}(f({\cal
  D}_A/\Lambda))$ can be expressed perturbatively
as~\cite{Chamseddine:2005zk,Chamseddine:2008zj}
\be\label{asymp-exp} {\rm Tr}(f({\cal D}_A/\Lambda))\sim
2\Lambda^4f_4a_0+2\Lambda^2f_2a_2+f_0a_4+\cdots
+\Lambda^{-2k}f_{-2k}a_{4+2k}+\cdots~.  \ee
Since the Taylor expansion of the $f$ function vanishes at zero, the
asymptotic expansion of the spectral action, in terms of the
geometrical Seeley-De Witt coefficients $a_n$, reduces to
\be \label{asympt} {\rm Tr}(f({\cal D}_A/\Lambda))\sim
2\Lambda^4f_4a_0+2\Lambda^2f_2a_2+f_0a_4~.  \ee
Hence, the cutoff function $f$ plays a r\^ole only through its momenta
$f_0, f_2, f_4$, three real parameters, related to the coupling
constants at unification, the gravitational constant, and the
cosmological constant, respectively. This action has to be considered
as the bare action at unification scale; to make extrapolations to
lower energy scales one has to use renormalisation group equations and
consider nonperturbative effects in the NCSG action.

The noncommutative spectral geometry model offers~\cite{ccm} a purely
geometric approach to the SM of particle physics, where the fermions
provide the Hilbert space of a spectral triple for the algebra and the
bosons are obtained through inner fluctuations of the Dirac operator
of the product ${\cal M}\times {\cal F}$ geometry. The model is in
agreement with particle physics data, such as the top quark
mass~\cite{ccm} and, as recent studies have
shown~\cite{cchiggs,Devastato:2013oqa}, it is also consistent with the
Higgs mass. It is worth noting that in the original model~\cite{ccm},
the Higgs mass in the zeroth order approximation was found to be 170
GeV, inconsistent with recent particle physics experiments. However,
in the original approach, the real scalar singlet, associated with the
Majorana mass of the right-handed neutrino, was integrated out and
replaced by its vacuum expectation value.  It was then
shown~\cite{cchiggs} that this singlet, whose presence has been also
argued in Ref.~\cite{Devastato:2013oqa}, is non-trivially mixed with
the Higgs doublet. This results to their masses being shifted,
rendering the model consistent with a 125 GeV Higgs and a 170 GeV top
quark. Finally, let us note that extensions to the Pati-Salam model
have been considered more recently~\cite{Chamseddine:2013rta}.

The NCSG model lives by construction at the Grand Unified Theories
(GUTs) scale --- the cutoff scale $\Lambda$ is set at the unification
scale --- offering a natural framework to study early universe
cosmology~\cite{Nelson:2008uy}-\cite{Sakellariadou:2012jz}.  The
gravitational part of the asymptotic expression for the bosonic sector of
the NCSG action, including the coupling between the Higgs field $\phi$
and the Ricci curvature scalar $R$, in Lorentzian signature ---
obtained through Wick rotation in imaginary time ---reads~\cite{ccm}
\begin{equation}\label{eq:0}
S_{\rm grav}^{\rm L}= \int \left(\frac{R}{2\kappa_0^2} + \alpha_0
C_{\mu\nu\rho\sigma}C^{\mu\nu\rho\sigma} + \tau_0 R^\star
R^\star - \xi_0 R|{\bf H}|^2 \right)
\sqrt{-g} {\rm d}^4 x\,;
\end{equation}
${\bf H}=(\sqrt{af_0}/\pi)\phi$, with $a$ a parameter related to
fermion and lepton masses and lepton mixing. At unification scale (set
up by $\Lambda$), $\alpha_0=-3f_0/(10\pi^2)$.

At this point one may wonder whether the quadratic
curvature terms in the action functional indicate the emergence of
negative energy massive graviton modes~\cite{stelle}. We will briefly highlight that
this is not the case. The higher derivative terms that are quadratic
in curvature lead to~\cite{Donoghue:1994dn}
\be \int\left ({1\over 2\eta} C_{\mu\nu\rho\sigma}C^{\mu\nu\rho\sigma}
-{\omega\over 3\eta} R^2 +{\theta\over \eta}E \right) \sqrt{-g}d^4x~;
\nonumber\ee
$E=R^\star R^\star$ denotes the topological term which is the
integrand in the Euler characteristic
\be
\int E\sqrt{-g}d^4x=\int R^\star R^\star \sqrt{-g}d^4x~.
\nonumber
\ee
The running of the coefficients $\eta, \omega, \theta$ of the higher
derivative terms is determined by the renormalisation group
equations~\cite{Donoghue:1994dn}.  The coefficient $\eta$ goes slowly
to zero in the infrared limit, so that $1/\eta={\cal O}(1)$ up to
scales of the order of the size of the universe. Note that $\eta(t)$ varies by at
most one order of magnitude between the Planck scale and infrared
energies.  All three coefficients $\eta(t), \omega(t), \theta(t)$ run to a
singularity at a very high energy scale ${\cal O}(10^{23}) {\rm GeV}$ (i.e., above the Planck scale).
To avoid low energy constraints, the coefficients of the quadratic
curvature terms $R_{\mu\nu}R^{\mu\nu}$ and $R^2$ should not exceed
$10^{74}$~\cite{Donoghue:1994dn}, which is indeed the case for the
running of these coefficients.

For simplicity and since it
will not influence our results, in what follows we neglect the
nonminimal coupling between the Higgs field and the Ricci
curvature. The NCSG equations of motion are~\cite{Nelson:2008uy}
\begin{equation}\label{2}
    G^{\mu\nu}_{(\mbox{NCSG})} = \kappa^2 T^{\mu\nu}_{({\rm matter})}\,,
\end{equation}
where $\kappa^2\equiv 8\pi G$ and
\begin{equation}\label{G-NCG}
  G^{\mu\nu}_{(\mbox{NCSG})} \equiv
  G^{\mu\nu}+\frac{1}{\beta^2}[2\nabla_\lambda \nabla\kappa
  C^{\mu\nu\lambda\kappa}+C^{\mu\lambda\nu\kappa}R_{\lambda\kappa}]\,;
\end{equation}
$G^{\mu\nu}$ is the (zeroth order) Einstein tensor, $T^{\mu\nu}_{\rm
matter}$ the energy-momentum tensor of matter and $\beta^{2}
=\displaystyle{5\pi^2/(6\kappa^2f_0)}$.\\
Using the Bianchi identity
$\nabla^\sigma R_{\mu\lambda\nu\sigma}=-\nabla_\lambda
R_{\mu\nu}+\nabla_\mu R_{\lambda\nu}$ and $2\nabla^\sigma R_{\lambda
\sigma}= \nabla_\lambda R$, the second term above reads
\begin{equation}\label{2C+CR}
2\nabla^\lambda \nabla^\sigma
C_{\mu\lambda\nu\sigma}+C_{\lambda\mu\sigma\nu}R^{\lambda\sigma}=
-\Box \left(R_{\mu\nu}-\frac{1}{6}g_{\mu\nu}R\right)
+\frac{1}{3}\nabla_\mu\nabla_\nu R
\end{equation}
 \[
-2R_{\mu\rho}R_\nu^{\,\,\,\rho}+\frac{2}{3}R\,
R_{\mu\nu}+\frac{1}{2}g_{\mu\nu}
\left(R_{\alpha\beta}R^{\alpha\beta}-\frac{R^2}{3}\right)\,,
 \]
where $\Box \equiv \nabla_\mu \nabla^\mu$.

The aim of this paper is to constrain the parameter $\beta$, which
corresponds to a restriction on the particle physics at unification
by making use of recent results obtained from Gravity Probe B
satellite, and then to improve this constrain by using results from torsion
balance experiments.  We will thus extend previous
studies~\cite{Nelson:2010rt, Nelson:2010ru} of one of us and
collaborators, where by using recent observations of pulsar timing, we
were able to set $\beta \geq 7.55\times 10^{-13} {\rm m}^{-1}$.

It is worth noting that one cannot constrain the other two free
parameters, namely $f_2, f_4$ unless one makes a (unjustified to our
opinion) {\sl ansatz} on how the coefficients of the terms appearing in the
action functional run with energy in the renormalisation group
equations.

\section{Gravitational Waves in Noncommutative Spectral Geometry}

\label{linear}
Neglecting the nonminimal coupling between the Higgs field and the
Ricci curvature, NCSG does not lead to corrections for homogeneous and
isotropic cosmologies~\cite{Nelson:2008uy}.  We will therefore
consider linear perturbations
\be
g_{\mu\nu}=\eta_{\mu\nu}+\gamma_{\mu\nu}~,\ee
around a Minkowski
background metric $\eta_{\mu\nu}$, so that to first order $g^{\mu\nu}=
\eta^{\mu\nu}-\gamma^{\mu\nu}$.  Considering the weak field
approximation we will be able to get analytically a lower bound on
$f_0$.\\
Defining
\be
\Box_\eta \equiv \partial_\rho \partial^\rho~~,~~
{\bar \gamma}_{\mu\nu}
\equiv\gamma_{\mu\nu}-\frac{1}{2}\eta_{\mu\nu}\gamma~,\ee
with
\be
\gamma =
\gamma^\mu_{\,\,\,\mu}=\eta^{\mu\nu}\gamma_{\mu\nu}~,\ee
the
$G^{\mu\nu}_{(\mbox{NCSG})}$ is corrected by higher order
contributions.\\
In the Lorentz (synchronous) gauge $\partial_\mu {\bar
\gamma}^{\mu\nu}=0$, it hence assumes the form~\cite{Nelson:2010rt}
%
\begin{equation}\label{G-NCG-gamma}
 G^{\mu\nu}_{(\mbox{NCSG})} =\nonumber\\ -\frac{1}{2} \Box_\eta{\bar
 \gamma}^{\mu\nu} +\frac{1}{2\beta^2} \left[\Box_\eta{\bar
 \gamma}^{\mu\nu}+\frac{1}{3}\left(\eta^{\mu\nu}\Box_\eta-\partial^\mu
 \partial^\nu\right)\gamma\right]~.
 \end{equation}
%
Introducing the tensor~\cite{Nelson:2010rt}
\begin{equation}\label{bar-h}
    {\bar h}_{\mu\nu}={\bar \gamma}_{\mu\nu}-\frac{1}{3\beta^2}\,
    {\cal
      Q}^{-1}\left(\eta_{\mu\nu}\Box_\eta-\partial_\mu\partial_\nu\right)\gamma\,,
\end{equation}
with~\footnote{We assume that the operator ${\cal Q}$ can be inverted since the terms in the r.h.s. of Eq.~(\ref{bar-h}) are well-defined and we are considering linear perturbations in the weak field approximation.}
\be
\label{def-Q}
{\cal Q}\equiv 1-\frac{1}{\beta^2}\, \Box_\eta~,
\ee
the trace of ${\bar h}_{\mu\nu}$ reads~\cite{Nelson:2010rt}
\be\label{trace-barWe define -h} {\bar h} =
-\left(1+\frac{1}{\beta^2}\, {\cal Q}^{-1}\Box_\eta\right)\gamma =-
{\cal Q}^{-1}\gamma\,. \ee
In terms of ${\bar h}^{\mu\nu}$ the linearised NCSG equation of
motion is~\cite{Nelson:2010rt}
\begin{equation}\label{waveeq}
    \left(1-\frac{1}{\beta^2}\Box_\eta \right)\Box_\eta {\bar
    h}^{\mu\nu}= - 2\kappa^2 T^{\mu\nu}_{\rm matter}\,;
\end{equation}
$T^{\mu\nu}_{\rm matter}$ is taken to lowest order in
$\gamma^{\mu\nu}$, so that it is independent of $\gamma^{\mu\nu}$ and
satisfies the conservation equation $\partial_\mu T^{\mu\nu}_{({\rm
    matter})}=0$.
We restrict to $\alpha_0 <0$ for Minkowski to be a
stable vacuum of the theory~\cite{Nelson:2010rt}, implying
$\beta^2>0$.
\\
Let us define the tensor
 \begin{equation}
  \chi^{\mu\nu} \equiv -\frac{1}{\beta^2}\Box_\eta {\bar h}^{\mu\nu}\,,
  \end{equation}
with $\chi_{\mu\nu}$ satisfying the
\begin{equation}\label{waveeqchi}
    (\Box_\eta-\beta^2)\chi^{\mu\nu} = - 2\kappa^2 T^{\mu\nu}_{({\rm matter})}\,.
\end{equation}
We denote by ${\gamma}^{\mu\nu}_{({\rm GR})}$  the Einstein's
theory of General Relativity (GR) metric. It fulfills the equation
\begin{equation}\label{BoxGammaE}
    \Box_\eta {\gamma}^{\mu\nu}_{ ({\rm
        GR})}=-2\kappa^2\left(T^{\mu\nu}_{(\rm matter)}-\frac{1}{2}\eta^{\mu\nu}T_{(\rm matter)}\right)\,,
\end{equation}
where $T_{({\rm matter})}$ is the trace of the energy-momentum tensor,
with solution
 \begin{equation}\label{gammaE-T}
    \gamma_{\mu\nu ({\rm GR})}({\bf r})=-2\kappa^2 \int d^3 {\bf r}^\prime
 \frac{{\cal T}_{\mu\nu}({\bf r}')}  {|{\bf r}-{\bf r}^\prime|}\,,
 \end{equation}
where
\be
{\cal T}_{\mu\nu}({\bf r}')\equiv \displaystyle{T_{\mu\nu({\rm
matter})}({\bf r}^\prime)-\frac{1}{2}\eta_{\mu\nu}T_{({\rm
matter})}({\bf r}^\prime)}~.
\ee
The linearised equation of motion, Eq.~(\ref{waveeq}), reads
\begin{equation}\label{waveeqh1}
    \Box_\eta ({\bar h}^{\mu\nu} + \chi^{\mu\nu})=-2\kappa^2
    T^{\mu\nu}_{({\rm matter})}\,,
\end{equation}
which has the same  form as the linearised equation for Einstein's
theory of GR in the synchronous gauge. 
Hence its solution can be written as
\begin{equation}\label{h=gamma+chi}
    {\bar h}^{\mu\nu}={\bar \gamma}^{\mu\nu}_{({\rm GR})} - \chi^{\mu\nu}\,.
\end{equation}
From Eqs.~(\ref{bar-h}), (\ref{h=gamma+chi}) and writing
$\gamma=-{\cal Q}({\bar \gamma}_{({\rm GR})}-\chi)$, with ${\bar
\gamma}_{({\rm GR})}=-{\gamma}_{({\rm GR})}$ and $\chi=\chi^\mu_\mu$,
we get
\begin{equation}\label{gammafinal}
    \gamma_{\mu\nu}=\gamma_{\mu\nu ({\rm GR})}-{\bar
      \chi}_{\mu\nu}+\phi_{\mu\nu}\,,
\end{equation}
where
 \begin{eqnarray}
   {\bar\chi}_{\mu\nu} &=&
   \chi_{\mu\nu}-\frac{1}{2}\eta_{\mu\nu}\chi\,, \label{bar-chi}\nonumber
   \\ \phi_{\mu\nu} &=& \left[\frac{1}{2}\eta_{\mu\nu}(1-{\cal
       Q})-\Omega_{\mu\nu} \right](\gamma_{({\rm
       GR})}-\chi)\,, \label{phi} \nonumber\\ & = & \frac{1}{3\beta^2}
   \, \Pi_{\mu\nu} (\gamma_{({\rm GR})}-\chi)\,,
\end{eqnarray}
with
\begin{eqnarray}
   \Pi_{\mu\nu} &\equiv& \frac{1}{2}\eta_{\mu\nu}\Box_\eta+
   \partial_\mu\partial_\nu \,, \nonumber\\ \Omega_{\mu\nu} & \equiv &
   \frac{1}{3\beta^2}(\eta_{\mu\nu}\Box_\eta-\partial_\mu\partial_\nu)\,. \label{Omega}
 \end{eqnarray}
In what follows we shall compute the terms in the right-hand-side of
Eq.~(\ref{gammafinal}).

Let us consider a rotating source producing a static gravitational
field, we then have
 \begin{equation}\label{tensorT}
    T^{\mu\nu}_{({\rm matter})}({\bf r})=-\rho({\bf r}) u^\mu u^\nu\,,
 \end{equation}
where $u^\mu$ is the four velocity ($u^\mu u_\mu=-1$) and $\rho=T_{({\rm
matter})}$ is the time independent matter density referred to the
frame rotating with the source. Setting the origin of coordinates at
the centre of mass, we get for large $r$:
 \begin{equation}\label{expansion}
    \frac{1}{|{\bf r}-{\bf r}^\prime|}= \frac{1}{r}+\frac{1}{r^3}\,
    \sum_{i=1}^3 x^i x^{\prime\, i}+\dots
 \end{equation}
where $r=|{\bf r}|$ and ${\bf r}=(x^1, x^2, x^3)$.
\\
Hence, we obtain the standard GR result (dipole approximation):
\bea\label{gammaEGM} \gamma_{00({\rm GR})}=\frac{2GM}{r}\,, \quad
\gamma_{ij({\rm GR})}=\frac{2GM}{r}\delta_{ij}\,,\nonumber\\
\gamma_{0i({\rm GR})}=\gamma_{i0({\rm GR})}=-\frac{4G}{r^3}({\bf
r}\wedge {\bf J})_i\,, \eea where
 \begin{equation}\label{defJ}
    M=\int \rho({\bf r}^\prime) d^3 {\bf r}^\prime \,, \quad {\bf
    J}=\int \rho({\bf r}^\prime)\, [{\bf r}^\prime \wedge {\bf v}]\,
    d^3 {\bf r}^\prime\,.
 \end{equation}
Note that we have neglected the kinetic energy term being of the second order
in the spatial velocity.
\\
Using the series expansion
 \begin{equation}\label{expansionexp}
    \frac{e^{-\beta|{\bf r}-{\bf r}^\prime|}}{|{\bf r}-{\bf
    r}^\prime|}\simeq e^{-\beta r}\left[\frac{1}{r}+\frac{1+\beta
    r}{r^3}\, \sum_{i=1}^3 x^i x^{\prime\, i}+\dots\right]\,,
 \end{equation}
the solution
 \begin{equation}\label{chi-T}
    \chi_{\mu\nu}({\bf r})=2\kappa^2 \int d^3 {\bf r}^\prime
    \frac{T_{\mu\nu({\rm matter)}}({\bf r}^\prime)}{|{\bf r}-{\bf
    r}^\prime|}\, e^{-\beta|{\bf r}-{\bf r}^\prime|}\, ,
 \end{equation}
with $\beta>0$, of Eq.~(\ref{waveeqchi}) can be explicitly written as
\bea\label{chicomponent} \chi_{00}=4GM\frac{e^{-\beta r}}{r}\,, \nonumber\\
\chi_{ij}\sim {\cal O}(v^i v^j)\sim 0\,,
\nonumber\\ \chi_{0i}=\chi_{i0}=-4G\frac{(1+\beta r)e^{-\beta
    r}}{r^3}\, ({\bf r}\wedge {\bf J})_i\,.  \eea
Hence, in the dipole approximation, we get \bea\label{barchicomponent}
{\bar \chi}_{00}=2GM\frac{e^{-\beta r}}{r}\,, \nonumber\\{\bar \chi}_{ij}=
2GM\frac{e^{-\beta r}}{r} \delta_{ij}\,,\nonumber\\ {\bar
  \chi}_{0i}={\bar \chi}_{i0}=-4G\frac{(1+\beta r)e^{-\beta r}}{r^3}\,
({\bf r}\wedge {\bf J})_i\,.  \eea
For a static gravitational field, the non-vanishing components of
$\phi_{\mu\nu}$ are $\phi_{00}$ and $\phi_{ij}$ given by:
 \bea\label{phi-ij} \phi_{00} &=& - \frac{2GM}{3}\frac{e^{-\beta
     r}}{r}\,, \nonumber \\ \phi_{ij} &=& - \frac{4GM}{3\beta^2
   r^3}\left\{\left[1+\left(1+\beta r-\frac{\beta^2 r^2}{2}\right)
   e^{-\beta r}\right]\delta_{ij} \right. \nonumber\\ &-&
 \left. 3\frac{x^i x^j}{r^2}\left[1+\left(1+\beta r+\frac{\beta^2
     r^2}{3}\right)e^{-\beta r}\right]\right\}\,, \eea
where we have used that
\be
\gamma_{({\rm GR})} - \chi = 4GM \left({1+e^{-\beta r}\over r}\right)~.
\ee
%
Introducing the metric potentials $\Phi, \Psi$ and the vector
potential ${\bf A}$, the metric reads
 \begin{equation}\label{elementline}
    ds^2 = -(1+2\Phi)dt^2+ 2{\bf A}\cdot d{\bf x} dt+(1+2\Psi) d{\bf x}^2 \,.
 \end{equation}
Assuming that for satellite orbits, the relation
\be
x^i x^j =
{r^2\over 3}\delta_{ij}
\ee
holds on the average, Eq.~(\ref{phi-ij}) simplifies to
\begin{equation}
    \phi_{ij}=\frac{10 G M}{9}\frac{e^{-\beta r}}{r}\,\delta_{ij}\,.
\end{equation}
In terms of $\Phi, \Psi, A_i$, the components of $\gamma_{\mu\nu}$ are
\begin{eqnarray}\label{gamma001}
    \gamma_{00}&=&-2\Phi=\frac{2GM}{r}\left(1-\frac{4}{3}e^{-\beta
    r}\right)\,, \nonumber \\ \gamma_{0i}&=&\gamma_{i0}= A_i \nonumber
    \\ &=&-\frac{4G}{r^3}[1-(1+\beta r)e^{-\beta r}]({\bf r}\wedge
    {\bf J})_i\,, \nonumber \\ \gamma_{ij} &=& 2\Psi \delta_{ij}
    \nonumber \\ &=& \frac{2GM}{r}\left[1+\frac{5}{9}e^{-\beta
    r}\right]\delta_{ij}\,,
 \end{eqnarray}
and the non-vanishing Christoffel symbols read
\bea\label{Christoffel} \Gamma^0_{0i}&=&\Gamma^i_{00}=-\partial_i
    \Phi\,, \nonumber\\ \Gamma^i_{0j}&=&\frac{1}{2}(\partial_i A_j-\partial_j
    A_i)\,, \nonumber\\ \Gamma^i_{jk}&=& \delta_{jk}\partial_i \Psi
    -\delta_{ij}\partial_k \Psi-\delta_{ik}\partial_j \Psi\,.  \eea
Notice that the modifications induced by the NCSG action to the
Newtonian potentials $\Phi$ and $\Psi$ as appear in
Eq.~(\ref{gamma001}) are similar to those induced by a fifth-force
through a potential~\cite{fischbach}
\be\label{fifthV}
V(r)=-\displaystyle{\frac{GMm}{r}\Big(1+\alpha e^{-r/\lambda}\Big)}~,
\ee
where $\alpha$ is a dimensionless strength parameter and $\lambda$ a
length scale.  In the following, we will put a lower bound on $\beta$,
or equivalently an upper bound on $\lambda$. We will then see that by using
current experimental data that constrain $\lambda$ we can set a stronger
constraint on the $\beta$ parameter of our model.

\section{Constraints from Gravity Probe B and from Torsion Balance}
The Gravity Probe B satellite contains a set of gyroscopes (in low
circular polar orbit with altitude $h=650$~km) that, according to
general relativity, will undergo a geodesic precession in the orbital
plane, as well as a Lense-Thirring --- frame-dragging --- precession
in the plane of the Earth equator.  The Lense-Thirring precession is
related to the off-diagonal components of the metric tensor of a
rotating gravitational source, so its experimental verification will
test the Einstein theory for gravitation.  The values (in units of
milliarcsec/year) of the geodesic precession and the Lense-Thirring
precession measured by the Gravity Probe B satellite and those
predicted by General Relativity are~\cite{GPBexp}
 \[
\begin{tabular}{ccc} \\ \hline
Effect & \ \ \ \ \ \ Measured & \ \ \ \ \ \ Predicted \\
  \hline Geodesic precession &\ \ \ \ \ \ $6602\pm 18$ &\ \ \ \ \ \
  6606 \\ \hline Lense-Thirring precession & $\ \ \ \ \ \ 37.2\pm 7.2$
  &\ \ \ \ \ \ 39.2 \\ \hline
\end{tabular}
 \]
Splitting the rate of an orbiting gyroscope precession into a part
generated by the metric potentials and one generated by the vector
potential, we get the following spin equation of motion for the gyro
spin three-vector ${\bf S}$~\cite{sch,Adler:1999yt}:
\begin{equation}\label{dSdt}
 \frac{d{\bf S}}{dt}=\frac{d{\bf S}}{dt}\Big|_{\rm G}+\frac{d{\bf S}}{dt}\Big|_{\rm LT}\,,
\end{equation}
where the instantaneous geodesic precession is
\begin{equation}
\label{dS-G}
\frac{d{\bf S}}{dt}\Big|_{\rm G}={\boldsymbol \Omega}_{\rm G}\wedge {\bf S}\ \ {\rm with}\ \ {\boldsymbol \Omega}_{\rm G}=\frac{1}{2}[\nabla(\Phi-2\Psi)]\wedge {\bf v}\,
\end{equation}
and the instantaneous Lense-Thirring precession is
\begin{equation}\label{dS-LT}
    \frac{d{\bf S}}{dt}\Big|_{\rm LT}={\boldsymbol \Omega}_{\rm LT}\wedge {\bf
    S}\ \ {\rm with}\ \  {\boldsymbol \Omega}_{\rm LT}=\frac{1}{2}\nabla\wedge
    {\bf A}\,.
\end{equation}
The geodesic and Lense-Thirring precession, ${\Omega}_{\rm G}$ and ${\Omega}_{\rm LT} $, respectively, can be written as the sum of two terms, one
obtained within GR and the other being the NCSG contribution. Thus,
\begin{eqnarray}\label{OmegaGfinal}
{\boldsymbol \Omega}_{\rm G}={\boldsymbol \Omega}_{\rm G (GR)}+{\boldsymbol \Omega}_{\rm G (NCG)}\,,
\end{eqnarray}
with
 \begin{eqnarray}\label{OmegaGE}
 {\boldsymbol \Omega}_{\rm G (GR)} &=& \frac{3GM}{2r^3}({\bf r}\wedge {\bf
   v})\,, \nonumber\\
   {\boldsymbol \Omega}_{\rm G (NCSG)} &=&
 -\frac{20}{27}(1+\beta r)\, e^{-\beta r}{\boldsymbol \Omega}_{\rm G (GR)}\,.
 \end{eqnarray}
Similarly,
\begin{equation}\label{OmegaLTfinal}
{\boldsymbol \Omega}_{\rm LT}={\boldsymbol \Omega}_{\rm LT (GR)}+{\boldsymbol \Omega}_{\rm LT (NCSG)}\,,
 \end{equation}
with
 \begin{eqnarray}\label{OmegaLTaverage}
    {\boldsymbol \Omega}_{\rm LT (GR)}& =& -\frac{2 G}{r^3}{\bf J}\,, \nonumber \\
    {\boldsymbol \Omega}_{\rm LT (NCSG)} &=& - e^{-\beta r}(1+\beta r +\beta^2
    r^2){\boldsymbol \Omega}_{\rm LT (GR)}\,,
 \end{eqnarray}
where we have assumed that on the average $\langle ({\bf J}\cdot
{\bf r}){\bf r}\rangle =0$.

We will use Eqs.~(\ref{OmegaGE}) and (\ref{OmegaLTaverage}) to
constrain the parameter $\beta$.  Here $r$ the sum of the Earth radius
$R_\oplus$ and the altitude $h$ of the satellite.  Setting the
geodesic precession $|{\boldsymbol \Omega}_{\rm G (GR)}|=6606$ mas/y and requiring
that $|{\boldsymbol \Omega}_{\rm G (NCSG)}| \lesssim |\delta {\boldsymbol \Omega}_{\rm G}|$, where
$|\delta {\boldsymbol \Omega}_{\rm G}| = 18$ mas/y, we get
 \begin{equation}\label{boundbetaG}
    \beta \gtrsim 10^{-6}\rm{m}^{-1}\,.
 \end{equation}
Note that we get the same lower bound for $\beta$ from the
Lense-Thirring precession, where $|{\boldsymbol \Omega}_{\rm LT (GR)}|=39.2$ mas/y and
$|{\boldsymbol \Omega}_{\rm LT (NCSG)}| \lesssim |\delta {\boldsymbol \Omega}_{\rm LT}|$ where
$|\delta {\boldsymbol \Omega}_{\rm LT}|=7.2$ mas/y.

The constraint on $\beta$, Eq.~(\ref{boundbetaG}) above, provides an
upper bound on $\lambda=\beta^{-1}$, namely $\lambda < 10^6{\rm
  m}$. It has been shown~\cite{kapner} that the inverse-square law
holds down to a length scale $\lambda=56\mu{\rm m}$ for $|\alpha|\leq
1$. Note that in our case $\alpha\sim {\cal O}(1)$, as one can easily
see from Eqs.~(\ref{gamma001}) and (\ref{fifthV}).

A more stringent constraint on $\beta$ can be obtained once we use
results from laboratory experiments design to test the fifth
force. Hence, by constraining $\lambda$ through torsion balance
experiments we will subsequently obtain a stronger lower bound to
$\beta$, or equivalently an upper bound to the momentum $f_0$ of the
cutoff function $f$.

The test masses have a typical size of $\sim 10$mm and their
separation is smaller than their size. As we have already mentioned
above, for our study $|\alpha| \sim {\cal O}(1)$, so that the tightest
constraint on $\lambda$ provided by E\"ot-Wash~\cite{eot} and
Irvine~\cite{irvine} experiments is~\cite{kapner}
\begin{equation}
 \lambda \lesssim 10^{-4}\mbox{m}\,,
 \end{equation}
implying
\begin{equation}\label{constr}
 \beta \gtrsim 10^4 \mbox{m}^{-1}\,.
 \end{equation}

\section{Conclusions}
In the context of NCSG we have studied the linearised field
equations in the limit of weak gravitational fields generated by a
rotating gravitational source. Then making use of recent experimental
data, we were able to constrain one of the free parameters of the
model, namely the moment of the cutoff function that is related to the
coupling constants at unification. First, we have studied the
precession of spin of a gyroscope orbiting about a rotating
gravitational source. Such a gravitational field gives rise, according
to General Relativity predictions, to the geodesic and the
Lense-Thirring precessions, the latter being strictly related to the
off-diagonal terms of the metric tensor generated by the rotation of
the source.  We have focused in particular on the gravitational field
generated by the Earth, and on the recent experimental results
obtained from the Gravity Probe B satellite, which has tested the
geodesic and Lense-Thirring spin precession with high precision. We
have calculated the corrections of the precession induced by NCSG
corrections. Requiring that the corrections are below the experimental
errors, we have inferred a lower bound on $\beta$, namely that $\beta
\gtrsim 10^{-6}$m$^{-1}$.  We then used laboratory fifth force
experiments to impose a more stringent constraint on the parameter
$\beta$; we thus obtained $\beta \gtrsim 10^4$m$^{-1}$.  Note that
this is a stronger constraint than the one
imposed~\cite{Nelson:2010ru} by studying the energy lost from binary
systems via emission of gravitational waves, and much stronger than
any constraint imposed so far to curvature squared terms (as for
instance in Ref.~\cite{stelle}).


\vspace{0.2in} {\bf Acknowledgments}:\\ It is a pleasure to thank Alain Connes for useful discussions. G.\ Lambiase thanks the ASI
(Agenzia Spaziale Italiana) for partial support through the contract
ASI number I/034/12/0.


\end{document}